\documentclass[english,showpacs,floatfix,amsmath,amssymb,12pt]{revtex4}
\usepackage[dvips]{graphicx}
\usepackage{longtable}
\usepackage{dcolumn}
\usepackage{latexsym}
\usepackage{babel}
\usepackage[latin1]{inputenc}
\usepackage{color}
\usepackage[colorlinks]{hyperref}
\newcommand\tabcaption{\def\@captype{table}\caption}
\def\al{\alpha}

\def\kp{\kappa}
\def\nb{\nabla}
\def\pa{\partial}
\def\vf{\varphi}

\def\om{\omega}

\def\la{\lambda}

\def\th{\theta}
\def\Th{\Theta}

\def\nn{\nonumber}

\def\ln{\mbox {ln}}
\def\wt{\widetilde}
\def\l{\left}
\def\r{\right}

\begin{document}
\title{\Large\bf The Series Solution to the Metric \\
of Stationary Vacuum with Axisymmetry}
\author{Ying-Qiu Gu}
\email{yqgu@fudan.edu.cn} \affiliation{School of Mathematical
Science, Fudan University, Shanghai 200433, China} \pacs{
04.20.Ha, 04.20.Cv, 04.20.-q}
\date{22nd March 2009}

\begin{abstract}
The multipole moments method is not only an aid to understand the
deformation of the space-time, but also an effective tool to solve
the approximate solutions of the Einstein field equation. However,
The usual multipole moments are recursively defined by a sequence
of symmetric and trace-free tensors, which are inconvenient for
practical resolution. In this paper, we develop a simple procedure
to generate the series solutions, and propose a method to identify
the free parameters by taking the Schwarzschild metric as a
standard ruler. Some well known examples are analyzed and compared
with the series solutions. \vskip .3cm\noindent {\bf Keywords:}
{\sl stationary metric, multipole moments, asymptotically flat}

\end{abstract}
\maketitle

\section{Intruduction}
\setcounter{equation}{0}

The asymptotically flat vacuum solutions with axisymmetry play an
important role in general relativity. To understand the global
structure of the metric caused by stars, we have the exact
solution families of Schwarzschild, Curzon and Kerr metrics and
some of their extensions to the electrovacuum solutions such as
Reissner-Nordstr\"om and Kerr-Newman metrics
\cite{exact1}-\cite{exact4}. After more than ten years of work,
Manko {\it et al}. got an axisymmetric solution with five free
parameters\cite{exact6}. This solution is suitable for a rapidly
rotating neutron star, in which the strong magnetic field and the
oblate shape should be taken into account\cite{exact7}.

When getting a realistic exact solution becomes more and more
difficult, the effective approximate method is a good alternation.
The multipole moments expansion was introduced to interpret the
physical meaning of the solutions to the Einstein equation. Geroch
defined a recursive relation of multipole moments for the static
asymptotically flat vacuum with axisymmetry by means of the
timelike Killing vector\cite{multi1,multi2}. The norm of the
Killing vector satisfies the Laplace equation in the 3-dimensional
hypersurface orthogonal to the Killing vector, which is reduced
from the Einstein equation. For the asymptotically flat
space-time, the norm of the Killing vector and the metric is
analytic near the spatial infinity, so the solution can be
expanded in series of $r^{-1}$ with multipole moments. The
multipole moments can be recursively defined by a sequence of
symmetric and trace-free tensors. This idea was generalized to the
stationary case by Hansen in \cite{multi3}, where two different
sets of multipole moments are employed. One is the mass moments
and the other is the angular momentum moments. The two sets of
moments can be generated from two potential functions defined by
the norm and the twist of the time-like Killing vector.

The formalism proposed in \cite{multi2,multi3} is in the covariant
form. Some theorems are proved based on this formalism. In
\cite{proof1}, Xanthopoulos proved that a stationary space-time is
static if and only if all of its current moments vanish, and
static space-time is flat if and only if all of its mass moments
vanish. Beig-Simon \cite{proof2} and Kundu\cite{proof3} proved
that, two spacetimes with the same multipole moments have the same
space-time geometry at large radii, where the multipole expansion
of the metric converges. This formalism was realized in the
Weyl-Lewis-Papapetrou  canonical metric\cite{multi0,
multi4,multi5}.

An another formalism of multipole moments for realistic
calculation of slowly changing systems or precisely stationary
ones was reviewed and developed by Thorne in \cite{multi6}. This
formalism is more manifest for practical resolution, which is
associated with a special kind of coordinate system called
`Asymptotical Cartesian and Mass Centered'(ACMC) coordinate
system. In such ACMC-$N$ coordinate system, the metric can be
expressed by multipole moments expansion, and the coefficients of
$r^{-(l+1)}(l\le N)$ terms are linearly combinations of the
spherical harmonics of order $l, l - 1 ,\cdots, 0$. The de Donder
coordinate system belongs to ACMC-$\infty$. The comparison and the
relation between the two formalisms were discussed in \cite{comp}.
This formalism in de Donder coordinate system and quotient
harmonic one was developed into an effective tool for solving the
approximate global solutions to the Einstein
equation\cite{appr1}-\cite{appr3}.

However, most of the previous works concerned mainly theoretical
aspect of the multipole moments, which are very inconvenient for
practical resolution of the Einstein's field equation. Besides,
the interpretation of the physical meanings of the parameters is
still a general problem in general relativity\cite{intp1, intp2},
although the primary motivation of Geroch to introduce the
multipole moments was to clarify this problem.

In this paper, we consider how to effectively solve the series
solutions to the stationary, asymptotically flat vacuum with
axisymmetry. This case is of important physical significant. At
first, we show that the canonical form of Weyl-Lewis-Papapetrou
coordinate system is an ACMC-$\infty$ one. Then we solve the
series solution straightforwardly in this coordinate system. To
understand the meanings of the free parameters in the solution, we
set the Schwarzschild solution as a standard ruler. By comparison
with this solution, we find that the bigger the absolute value of
the dimensionless free coefficients, the larger the deformation
and convection of the gravitating source. Noticing its generality,
the series solution is an effective alternation to the exact one,
and is helpful to understand the abundant structure of the
space-time.

\section{The series solution in Weyl-Lewis-Papapetrou metric}
\setcounter{equation}{0}

To describe the axisymmetric space-time, the canonical form of the
Weyl-Lewis-Papapetrou metric is the simplest one\cite{exact1}. The
line element is equivalent to
\begin{eqnarray}
ds^2=U (d t + W d\vf)^{~2} -V(d\rho^2+dz^2)-U^{-1}\rho^2 d\vf^2,
\label{1.1}
\end{eqnarray}
where $(t,\rho,z,\vf)$ is the coordinates of the geometrical
meaning near the cylindrical coordinates in Minkowski space.
$(U,V,W)$ only depend on $(\rho,z)$. In this paper we take $G=c=1$
as units. Since the following calculations take the dimensionless
form, the units will not be confused. The canonical form
(\ref{1.1}) has an important property, that is,

{\bf Lemma 1}. {\em Except for the translation $t=\wt
t+t_0,\vf=\wt\vf +\vf_0$ related to two Killing vectors
$(\pa_t,\pa_\vf)$ and $z=\pm (\wt z+z_0)$, the form of metric
(\ref{1.1}) removes other uncertainty caused by coordinate
condition.}

This property can be checked as follows. The transformation
$(\rho,z)\to (\wt \rho,\wt z)$ keeping $V(d\rho^2+dz^2)=\wt V(d\wt
\rho^2+d\wt z^2)$ is a conformal transformation
\begin{eqnarray}
\rho=\Re [f(\wt\rho\pm \wt z i)],\qquad z=\Im [f(\wt\rho\pm \wt z
i)],
\end{eqnarray}
where $f$ is any given analytic function. But the restrictions
$\rho^2 d\vf^2=\wt\rho^2 d\vf^2$ and $\rho\ge0$ give $f=\wt\rho\pm
(\wt z+z_0)i$, then we have  $z=\pm (\wt z+z_0)$.

Except for the axial symmetry, the celestial body in equilibrium
and the related stationary metric still have another top-bottom
reflection symmetry, which means the metric will be the even
functions of $z$ by setting the origin $(\rho=0,z=0)$ at the mass
center. In what follows we only discuss this case, that is,
$(U,V,W)$ are even functions of $z$, although the discussion is
also valid in the case including odd terms of $z$. So if we set
the direction of $z$, the relation between coordinate and metric
is completely fixed.

The form (\ref{1.1}) is not convenient for the following
calculation, so we make a polar coordinate transformation
\begin{eqnarray}
\rho=r\sin\th,\qquad z=r\cos\th, \label{1.2}
\end{eqnarray}
then the line element (\ref{1.1}) becomes
\begin{eqnarray}
ds^2=U (d t + W d\vf)^{~2} -V(dr^2+r^2d\th^2)-U^{-1} r^2\sin^2\th
d\vf^2. \label{1.3}
\end{eqnarray}
Calculating the Ricci tensor $R_{\mu\nu}$, we get the independent
components among the Einstein equations for the axisymmetric
vacuum\cite{exact1}
\begin{eqnarray}
\pa_r^2U+\frac 2 r \pa_rU + \frac {\cot\th} {r^2} \pa_\th U +
\frac 1 {r^2} \pa_\th^2 U -\frac {|\nb U|^2} U  + \frac {U^3|\nb
W|^2}{r^2\sin^2\th}
&\equiv&-2V R_{tt}=0, \label{1.4} \\
\pa_r^2 W -\frac {\cot\th} {r^2} \pa_\th W +\frac 1 {r^2}
\pa_\th^2 W +\frac 2 U(\nb U\cdot\nb W) &\equiv& \frac {2V}
U(WR_{tt}+R_{t\vf})=0,~~ \label{1.5}
\end{eqnarray}
where $\nb =\l(\pa_r, r^{-1}\pa_\th\r)$. And then $V$ can be
integrated from
\begin{eqnarray}
\frac {\pa_r V}V +\frac {\pa_r U}U - \frac {r\sin^2\th}{2U^2}\l(
(\pa_rU)^2 +\frac {2\cot\th} r \pa_r U \pa_\th U \r. -\l. \frac 1
{r^2}(\pa_\th U)^2 \r) &+& \nn \\
\frac {U^2} {2r} \l((\pa_r
W)^2 +\frac {\cot\th} r \pa_r W\pa_\th W \r.-\l. \frac 1 {r^2}(\pa_\th W)^2 \r) &\equiv& \nn \\
-\frac 1 r \sin^2\th ( r^2 R_{rr}-R_{\th\th}) - 2\cos\th\sin\th
R_{r\th}
&=& 0,\label{1.6}\\
\frac {\pa_\th V}V +\frac {\pa_\th U}U + \frac
{r^2\sin^2\th}{2U^2}\l( \cot\th [(\pa_rU)^2-\frac 1 {r^2}(\pa_\th
U)^2 ] -\frac {2} r \pa_r U \pa_\th U \r)
 &-& \nn \\
  \frac {U^2} {2} \l(\cot\th [(\pa_rW)^2-\frac 1 {r^2}(\pa_\th
W)^2 ] - \frac 1 r \pa_r W \pa_\th W  \r) &\equiv& \nn \\
\cos\th\sin\th (r^2 R_{rr}-R_{\th\th}) -2r\sin^2\th R_{r\th} &=&
0. \label{1.7}
\end{eqnarray}
Substituting $R_{\mu\nu}=\kp (T_{\mu\nu}-\frac 1 2
g_{\mu\nu}T^\al_\al)$ into the above equations, we get the
dynamical equations for interior metric, which can also be solved
by series.

The solutions to (\ref{1.4})-(\ref{1.7}) of most physical
interests are the asymptotically  flat cases, that is, the
solution satisfies the boundary conditions at $r\to \infty$
\begin{eqnarray}
U\to 1-\frac {2m} r+O(r^{-2}),\quad W \to \frac {4 J}
r\sin^2\th+O(r^{-2}),\quad V\to 1+\frac {2m} r+O(r^{-2}),
\label{1.8}
\end{eqnarray}
in which $m$ is the total gravitational mass, and $J$ the angular
momentum of the source.  (\ref{1.8}) means $r=\infty$ is the
regular point of the solution. On the other hand, the main part of
derivatives in (\ref{1.4}) and (\ref{1.5}) are the linear elliptic
operators. So the vacuum solutions $(U,W)$ are analytic functions
in the neighborhood of $r=\infty$\cite{proof2,proof3}, and can be
expressed as Taylor series of $r^{-1}$, then we generally have
\begin{eqnarray}
U=1-\frac {R_s} r +\frac 1 {r^2} \sum_{n=0}^\infty \frac {\wt A_n}
{r^{n}},\qquad W=\sum_{n=1}^\infty \frac {\wt B_n} {r^{n}},
\label{1.9}
\end{eqnarray}
where $(R_s, \wt A_n, \wt B_n)$ are functions of $\th$ to be
determined. For the regular metric, $(U,V)$ are bounded even
functions of $z$, or equivalently, $(R_s, \wt A_n, \wt B_n)$ are
bounded functions of $\cos^2\th$.

Substituting (\ref{1.9}) into (\ref{1.4}), and then expand it into
Taylor series of $r^{-1}$,  we can get the equations for the
coefficients
\begin{eqnarray}
R_s''+\cot\th R_s'&=&0,\label{1.10}\\
\wt A_0''+\cot\th \wt A_0'+2 \wt A_0&=&(R_s')^2+R_s^2.
\label{1.11}
\end{eqnarray}
The bounded solution of reflecting symmetry with respect to $z=0$
or $\th=\frac \pi 2$, we call it {\bf regular solution} in what
follows, is given by
\begin{eqnarray}
R_s=2m,\qquad \wt A_0=\frac 1 2 R_s^2, \label{1.12}
\end{eqnarray}
where $m$ is a constant, which means the total mass of the source.
$m=0$ corresponds to the Papapetrou class of solutions or flat
space-time. The case of $m<0$ is unphysical. So only the case with
positive mass $m>0$ has the most physical interest, and we only
discuss this case in what follows. Obviously, $R_s$ is the
corresponding Schwarzschild radius with dimension of length.

By virtue of the excellent structure of (\ref{1.4}) and
(\ref{1.5}), we have the following results.

{\bf Theorem 2}. {\em For the regular solution (\ref{1.9}) with
starting value (\ref{1.12}), it takes the following form
\begin{eqnarray}
U&=&1-\frac {R_s} r +\frac 1 2  \l(\frac {R_s} {r}\r)^2 +\l(\frac
{R_s} {r}\r)^2 \sum_{n=1}^\infty { \wt A_n} {\l(\frac {R_s}
r\r)^{n}},\label{1.13}\\
W&=&R_s \sum_{n=1}^\infty { \wt B_n} \l(\frac {R_s} r\r)^{n},
\label{1.14}
\end{eqnarray}
in which
\begin{eqnarray}
\wt A_n = \sum_{k=0}^{N_n} \wt a_{nk} \cos(2k\th), \qquad \wt B_n
= \sum_{k=1}^{N_n} \wt b_{nk} [\cos(2k\th)-1], \label{1.15}
\end{eqnarray} where $(\wt a_{nk}, \wt b_{nk})$ are dimensionless constants to be determined, and
$N_n=\l[\frac {n+1} 2\r]$ stands for the integer part of $\frac 1
2 (n+1)$, namely, $N_{2k-1}=N_{2k}=k$.}

{\bf Proof}. Substituting (\ref{1.13}) and (\ref{1.14}) into
(\ref{1.4}) and (\ref{1.5}) and expand them in series, we get the
following equations
\begin{eqnarray}
\wt A_1''+\cot\th \wt A_1' +2\cdot3\wt A_1&=&-1,\label{1.17}\\
\wt A_2''+\cot\th \wt A_2' +3\cdot4\wt A_2&=&-\frac 1 2 -6\wt
A_1-\frac 1 {\sin^2\th}\l((\wt B_1')^2+\wt B_1^2\r),\label{1.18}\\
\wt B_1''-\cot\th \wt B_1' +1\cdot2\wt B_1&=&0,\label{1.19}\\
\wt B_2''-\cot\th \wt B_2'+2\cdot3 \wt B_2 &=&2 \wt B_1.
\label{1.20}
\end{eqnarray}
Solve the equations, we find that the regular solutions take the
forms (\ref{1.15}). We check the succeeding equations by
mathematical induction.

Assume the forms (\ref{1.15})  hold for $n=2N-1$ and $n=2N$. Since
the equations (\ref{1.4}) and (\ref{1.5}) are homogeneous with
respect to $r$, and the highest derivatives with respect to $r$ is
linear. Calculation shows that the equations for $n=2N+1$ and
$n=2N+2$ take the following form
\begin{eqnarray}
\wt A_{2N+1}''+\cot\th \wt A_{2N+1}' +({2N+2})({2N+3}) \wt A_{2N+1}&=&P_1-4(N+1)\wt A_{2N},\label{1.21}\\
\wt A_{2N+2}''+\cot\th \wt A_{2N+2}' +({2N+3})({2N+4}) \wt A_{2N+2} && \nn\\
-4\wt b_{11}[2\cot\th\wt B_{2N+1}'
+(2N+1)\wt B_{2N+1}]&=&P_2-2(2N+3)\wt A_{2N+1} ,\label{1.22}\\
\wt B_{2N+1}''-\cot\th \wt B_{2N+1}' +({2N+1})({2N+2}) \wt B_{2N+1}&=&Q_1+4N \wt B_{2N},\label{1.23}\\
\wt B_{2N+2}''-\cot\th \wt B_{2N+2}' +({2N+2})({2N+3}) \wt
B_{2N+2}&=&Q_2+2(2N+1)\wt B_{2N+1},\label{1.24}
\end{eqnarray}
where $(P_k,Q_k)$ are polynomials determined by functions $(\wt
A_{2j-1},\wt A_{2j},\wt B_{2j-1},\wt B_{2j})$,$(j\le N)$ and their
derivatives, which all take the forms (\ref{1.15}). Since
$\cos(2n\th)=\frac 1 2 (e^{2n\th i}+e^{-2n\th i})$, under the
constraint of the index of $\l(\frac {R_s} r\r)^n$, we find
\begin{eqnarray}
P_k = \sum_{n=0}^{N+1} p_{kn} \cos(2n\th), \qquad Q_k =
\sum_{n=0}^{N+1} q_{kn} \cos(2n\th), \label{1.25}
\end{eqnarray} with constants $(p_{kn}, q_{kn})$. Substituting
(\ref{1.25}) into (\ref{1.21})-(\ref{1.24}) and solving the
solutions with finite and symmetrical conditions, the solutions
$(\wt A_{2N+2},\wt B_{2N+1})$ have the forms (\ref{1.15}), but
$(\wt A_{2N+1},\wt B_{2N+2})$ will introduce extra terms similar
to the second kind Legendre functions, which include the following
factor
\begin{eqnarray}
F(\th)=\ln \l(\frac {1+\cos\th}{1-\cos\th}\r)\sin\th\cos\th.
\end{eqnarray}
$F(\th)$ also satisfies the boundary condition and symmetry.
Nevertheless, the terms including this factor will soon be
cancelled at the next step of resolution. Then we prove the
regular solutions always take the forms (\ref{1.15}).

(\ref{1.15}) means that the Weyl-Lewis-Papapetrou coordinate
system (\ref{1.1}) is an ACMC-$\infty$ one. (\ref{1.15}) can be
transformed into the Legendre polynomials, but the Legendre
polynomials are less convenient to calculate the nonlinear terms
than the trigonometric functions. If the solutions of $(U,W)$ is
determined, $V$ can be solved only by (\ref{1.6}) with boundary
condition $V(\infty,\th)=1$. Then the stationary axisymmetrical
metric of vacuum is determined in principle. We determine the
coefficients of the series in the next section.

\section{Coefficients of the series and examples}
\setcounter{equation}{0}

Similarly to determine the coefficients of the special functions
by recursion, the theorem 2 implies that, the resolution of
(\ref{1.4}) and (\ref{1.5}) can be transformed into a problem of
solving coefficients by recursive relation. For convenience of
calculation, we make transformation $U=\exp\l(\frac u 2\r)$, then
(\ref{1.4}) and (\ref{1.5}) become
\begin{eqnarray}
r^2 \pa_r^2u+2 r \pa_r u +  {\cot\th}  \pa_\th u + \pa_\th^2
u +\frac {2e^u}{r^2 \sin^2\th} |r^2(\pa_r W)^2+(\pa_\th W)^2| &=& 0,\label{2.1}\\
r^2 \pa_r^2 W - {\cot\th} \pa_\th W + \pa_\th^2 W + (r^2\pa_r u
\pa_r W+\pa_\th u \pa_\th W) &=& 0.\label{2.2}
\end{eqnarray}
and (\ref{1.13})-(\ref{1.15}) become
\begin{eqnarray}
u = \sum_{n=1}^\infty { A_n} {\l(\frac {R_s} r\r)^{n}},&&\qquad
W=R_s \sum_{n=1}^\infty { B_n} \l(\frac {R_s} r\r)^{n},
\label{2.3}\\
 A_n = \sum_{k=0}^{N_n-1} a_{nk} \cos(2k\th),&&\qquad
 B_n = \sum_{k=1}^{N_n} b_{nk} [\cos(2k\th)-1],
\label{2.4}
\end{eqnarray}
where $N_n=\l[\frac {n+1} 2\r]$.

Substituting (\ref{2.3}) and (\ref{2.4}) into the original
equation (\ref{2.1}) and (\ref{2.2}), and expanding them in
series, we get the relations among the parameters
$(a_{kn},b_{kn})$. (see appendix). In each term, the parameters
$(a_{kn},b_{kn})$ with the larger indexes always take linear form
with bigger coefficients, which correspond to the linear
Laplace-like operator in (\ref{2.1}) and (\ref{2.2}). So it is
easy to solve the coefficients. However, the equation are
underdetermined for the parameters $(a_{kn},b_{kn})$, and we have
the following two sequences of free parameters. Let
\begin{eqnarray}
&& a_{10} = -2,\quad a_{31} = m_1,\quad a_{52} = m_2,\quad
a_{2k+1,k} =
m_k, \label{2.7}\\
&& b_{11} = -\frac 1 2 w_1,\quad b_{32} = -\frac 1 4 w_2,\quad
b_{2k-1,k} = -\frac 1{2k}w_k,~~(k=1,2,3,\cdots). \label{2.8}
\end{eqnarray} where $\{m_k,w_k\}$ are dimensionless numbers
determined by the multipole moments of the energy-momentum tensor
distribution of the source, $a_{10} = -2$ is determined by
(\ref{1.8}), and the factor $\frac 1{2k}$ is introduced to scale
the coefficients in the recursive relation. The free parameters
(\ref{2.7}) and (\ref{2.8}) correspond to `null data' in
papers\cite{convg1,convg4},  but the definitions are different.
The coefficients can be recursively expressed as the polynomials
of the free parameters(see appendix). Solving the coefficients, we
get the series solutions of the metric
\begin{eqnarray}
u &=& -2 \frac {R_s}r + \frac 1 3 m_1 ( 1 +3\cos 2\th ) \l(\frac
{R_s}r\r)^3 -\frac 1 2 w_1^{2} ( 1+\cos 2 \th ) \l(\frac
{R_s}r\r)^4 +\nn\\
&& [m_2\cos 4 \th + ( \frac 2 7 {w_1}^{2}+\frac 4 7 m_{2} ) \cos 2
\th + {\frac {8}{35}} { w_1}^{2}+{\frac {9}{35}} m_{2} ] \l(\frac
{R_s}r\r)^5 + O(r^{-6}),\label{2.10*}\\
U&=&1-\frac {R_s}r+\frac 1 2 \l(\frac {R_s}r\r)^2+  ( \frac 1 2
m_1\cos  2 \theta  +\frac 1 6 m_{1} -\frac 1 6 ) \l(\frac
{R_s}r\r)^3+\nn\\
&& [ ( -\frac 1 4 { w_1}^{2}-\frac 1 2 m_1 ) \cos 2 \th +\frac
1{24}- \frac 1 4 {w_1}^{2}-\frac 1 6 m_1 ] \l(\frac
{R_s}r\r)^4+O(r^{-5}),\label{2.10}\\
W &=&{ R_s} \sin^2 \th \l\{ {w_1 \frac {{ R_s}}{r}}+\frac {w_1} 2
\l(\frac {R_s}r\r)^2+( w_2\cos  2 \th +\frac 3 5 w_2+\frac 1 5 w_1
) \l(\frac
{R_s}r\r)^3+\r. \nn\\
&&\l. [
 (\frac  3 4 w_2+\frac 1 8 m_1 w_1 ) \cos  2 \th +\frac 1 {15}
 w_1+{\frac {5}{24}} m_1 w_1+{\frac {9}{20}
} w_2 ]  \l(\frac {R_s}r\r)^4 +O(r^{-5}) \r\},\label{2.12}\\
V &=& 1+\frac {R_s}{r}+ (\frac 1 8 \cos 2 \th+\frac 3 8 ) \l(\frac
{R_s}r\r)^2+ [ ( -\frac 1 2 m_1+\frac 1 8 ) \cos 2 \th -\frac 1 6
m_1+\frac 1 {24} ]  \l(\frac
{R_s}r\r)^3+\nn\\
&& [( {\frac {9}{64}} { w_1}^{2}-{\frac {5}{32}} m_1+{\frac
{1}{256}} ) \cos
 4 \th + ( -\frac 3 8 m_1+\frac 3 {16} {w_1}^{2}+{\frac {3}{ 64}} )
\cos  2 \th +\nn\\
&& {\frac {11}{64}} {w_1}^{2}-{\frac {7}{ 768}}-{\frac {13}{96}}
m_1 ]  \l(\frac {R_s}r\r)^4  +O(r^{-5}) . \label{2.11}
\end{eqnarray}
Comparing (\ref{2.10})-(\ref{2.12}) with (\ref{1.8}), we learn the
physical meaning of $R_s$ and $w_1$,
\begin{eqnarray}
R_s= 2m,\qquad J = \frac 1 4 w_1 R_s^2= w_1 m^2.\label{2.13}
\end{eqnarray}
That is, $R_s$ corresponds to the Schwarzschild radius, $w_1$ to
the angular momentum.

Now we compare the series solution with some exact metrics, i.e.
the Curzon, Schwarzschild and Kerr solutions. The Curzon metric is
diagonal,
\begin{eqnarray}
U=\exp\l(-\frac{R_s} r \r ),\quad V=\exp\l(\frac {R_s}{r}- \frac 1
4 \l(\frac{R_s} r\r)^2 \sin^2\th \r),\quad W=0. \label{2.14}
\end{eqnarray}
Expanding (\ref{2.14}) into Taylor series and then comparing them
with (\ref{2.10})-(\ref{2.12}), we find that all free parameters
vanish
\begin{eqnarray}
m_k=w_k\equiv 0,\qquad (\forall k). \label{2.15}
\end{eqnarray}
Evidently, such results is caused by the coordinate system.

For the Kerr and Schwarzschild solution, under some parameter
transformation, in the canonical coordinate system (\ref{1.3}) it
becomes
\begin{eqnarray}
U  &=& 1-\frac {2({ R_+ } + { R_- } + R_s \cosh \om )R_s\cosh\om}{
[ ( { R_+ }-{ R_- } ) ^{2}+{{ R_s}}^{2} ] \cosh^2 \omega +2  ( {
R_+ }+{ R_- } ) {
 R_s} \cosh \omega  +4 { R_+ }  R_- } ,\label{2.16}
\\
V & =& 1+{\frac { [
 ( { R_+}-{ R_-} )^{2}+R_s ^{2} ] \cosh^2 \omega + 2( R_+ + R_-) {
  R_s }\cosh \omega  }{4 R_+ R_-}},\label{2.17}
\\
W  &=& {\frac {
 [  ( { R_+ }-{ R_- } ) ^{2}- R_s^{2} ]
 ( { R_+ }+{ R_- }+{ R_s } \cosh \omega)\sinh \omega }{[ ( { R_+ }-{ R_- } ) ^{2}-{{ R_s }}^{ 2}
]  \cosh^2 \omega +4  R_+ R_- }}, \label{2.18}
\end{eqnarray}
in which
\begin{eqnarray}
R_\pm=\sqrt {r^{2} \cosh^2 \omega \pm 2r m \cos\theta \cosh \omega
 +m^2},\quad R_s=2m, \quad \tanh \om = \frac {2J} {m^2}.
\end{eqnarray}
$\om=0$ corresponds to the Schwarzschild solution in the
Weyl-Lewis-Papapetrou coordinate system. The above formalism is
introduced for convenience of the following comparison. In
contrast the Taylor series of the above functions with
(\ref{2.10})-(\ref{2.12}), we get the following sequence of the
free parameters,
\begin{eqnarray}
m_1 &=& -\frac 1 8 + \frac 3 8 {\al }^{2},\qquad  m_2 = -{\frac {
7}{2^{9}}}+ {\frac {17  }{ 2^{8}}\al ^{2}}-{\frac
{5\cdot 7}{ 2^{9}}\al ^{4}} ,\nn \\
m_3 &=& -{\frac { 3\cdot 11 }{ 2^{14}}}+{\frac { 11\cdot 19 }{
 2^{14}} \al ^{2}}-{ \frac {  17 \cdot 23
}{2^{14}}\al ^{4}}+{\frac { 3 \cdot 7 \cdot 11 }{ 2^{14}}{ \al
}^{6}},
\nn \\
m_4&=&-{\frac {5 \cdot 11\cdot 13 }{  2^{21}}} +{\frac {
 3\cdot 5 \cdot 7 \cdot 13 }{ 2^{19}}\al ^{2}} -\frac {   3 \cdot
5^{3} \cdot 19
 }{  2^{20}} \al ^4+{\frac {59\cdot 67}{ 2 ^{19}}
\al ^{6}}-{\frac { 3^{2}\cdot 5 \cdot 11\cdot13}{  2^{21}} \al^8},~~~~\label{2.20}\\
w_1 &=& \frac 1 2 \al,\qquad w_2={\frac {3}{32}}\al -{\frac
{5}{32}}\al ^{3}\qquad w_3 = \frac{3^4}{2^{12}}\al -\frac {3\cdot
41}{2^{11}} \al ^3+\frac {3^3\cdot 7}{2^{12}}\al ^5,\nn \\
w_4 &=& \frac{11\cdot 13}{2^{15}}\al -\frac{13\cdot47}{2^{15}}\al
^3+\frac {881}{2^{15}}\al ^5-\frac{3\cdot 11\cdot 13}{2^{15}}\al
^7, ~~\cdots \nn
\end{eqnarray}
where $ \al =\tanh \om<1$. The sequences vanish quite fast.

$\al=0$ corresponds to the Schwarzschild solution. In this case,
the `mass moments' $m_k\ne 0$, which is caused by the coordinates.
So to understand these data we need a unified standard.

Evidently, the above calculation shows that any stationary metric
with axisymmetry is identical with two sequences $\{m_n,w_n\}$, in
which $\{m_n\}$ is mainly related to the multipole moments of mass
density, and $\{w_n\}$ to the current distribution. On the
contrary, for any given sequences $\{m_n,w_n\}$ with suitable
upper bounds, the series is a solution to the Einstein equation of
vacuum in the region of its convergence. So the relation between
the radius of  convergence of the series and the values of free
parameters $\{m_n,w_n\}$ is important. The convergence of the
multipole moments serieses were discussed in different context.
The conditions of convergence for the static solutions were
established in \cite{multi4, convg1,convg2}, and the conditions
for the stationary ones were given in \cite{convg3, convg4}.
Although the results ensure the convergence of the above series,
how to derive the concrete constraint for free parameters is quite
complicated.

In the static case, the radius of convergence may be directly
derived from the exact solution. The result is enlighten, so we
give some analysis. Since $W=0$, (\ref{2.1}) becomes Laplace
equation, and its general asymptotically flat solution is the Weyl
class(\cite{exact2},Ch.18),
\begin{eqnarray}
u=\sum_{n=0}^N m_{n} P_{n}(\cos\th)\l(\frac {R_s} r\r)^{n+1},\quad
(m_0=-2,~N\le\infty),\label{4.1}
\end{eqnarray}
where $P_n$ are the Legendre polynomials of $n$ degree. For any
$N$, the function $V$ can be also exactly solved from
(\ref{1.6})\cite{exact2}. The Curzon solution is the simplest case
with $N=0$. In the case $N=\infty$, we have

{\bf Theorem 3}. {\em For the series solution (\ref{4.1}) with
$N=\infty$, if the free parameters $m_n$ satisfy the following
condition
\begin{eqnarray}
|m_n| \le C (n+1)^K \la^n,  \label{4.2}
\end{eqnarray}
where $(C > 0,\la>0,K)$ are given numbers independent of $n$, then
the series solution (\ref{4.1}) and all its derivatives converge
in the region
\begin{eqnarray}
r>r_0 \equiv \la R_s.
\end{eqnarray}}

{\bf Proof}. By the property of the Legendre polynomials, for all
$n$ we have
\begin{eqnarray}
|P_n|&\le& 1 ,\label{4.4} \\
\frac d {d\th} P_n(\cos\th)&=&-(2n-1)\sin\th
P_{n-1}(\cos\th)+\frac d {d\th} P_{n-2}(\cos\th).\label{4.5}
\end{eqnarray}
By (\ref{4.5}) we get
\begin{eqnarray}
\l|\frac d {d\th} P_n(\cos\th)\r| &\le& (2n-1)+\l| \frac d {d\th}
P_{n-2}(\cos\th)\r| \nn\\
&\le& (2n-1)+(2n-5)+(2n-9)+\cdots  = \frac 1 2 n(n+1).\label{4.6}
\end{eqnarray}
Substituting (\ref{4.2}), (\ref{4.4}) and (\ref{4.6}) into
(\ref{4.1}), we find that the series and its derivatives are
controlled by,
\begin{eqnarray}
|u| &<& C \sum_{n=1}^\infty (n+1)^K \l(\frac {\la R_s} r\r)^{n+1}, \\
|\pa_r u| &<& \frac C { r }\sum_{n=1}^\infty (n+1)^{K+1} \l(\frac
{\la R_s}
r\r)^{n+1}, \\
|\pa_\th u| &<& \frac C 2 \sum_{n=1}^\infty (n+1)^{K+2} \l(\frac
{\la R_s} r\r)^{n+1}.
\end{eqnarray}
The right hand serieses  all converge in the region $r>r_0$, so
$(u,\pa_r u,\pa_\th u)$ are absolutely convergent in the region
$r>r_0$, and uniformly  absolutely convergent in the region $r\ge
r_1$ for any given $r_1>r_0$. Similarly we can check the results
for higher order derivatives. Integrating (\ref{1.6}), we find the
radius of convergence of $V$ is also $r_0$. The proof is finished.

In the stationary case with $W \ne 0$, the situation becomes more
complicated due to the nonlinear terms, because $(a_{nk}, b_{nk})$
are polynomials of $(m_j,w_j)$ of $j \le \left[ \frac{n}
{2}\right]$. However, from the recursive relations, we find the
following inequalities seems true,
\begin{eqnarray}
|a_{n k}|<C M_n (n+1)^K,\qquad |b_{nk}|<C M_n (n+1)^K, ~~\forall
(n,k), \label{4.10}
\end{eqnarray}
where $(K>0, C>0)$ are constants independent of $n$, and
\begin{eqnarray}
M_n=\max_{j\le \l[\frac n 2 \r]}\l\{|m_j|,|w_j|,|m_j|^{\frac n
2},|w_j|^{\frac n 2}\r\}.
\end{eqnarray}
In principle, the relation (\ref{4.10}) can be derived from the
results in \cite{convg3, convg4}. If (\ref{4.10}) holds, the
series solutions (\ref{1.13})-(\ref{1.15}) will be controlled by
\begin{eqnarray}
|U| < C_0 \sum_{n=0}^{\infty} (n+1)^{K+1}\l(\frac {\la R_s}r\r)^n
, \quad |W| < C_1 \sum_{n=0}^{\infty} (n+1)^{K+1}\l(\frac {\la
R_s}r\r)^{n+1}, \label{4.14}
\end{eqnarray}
where $(C_0,C_1)$ are constants independent of $n$, and
\begin{eqnarray}
\la = \lim_{n\to\infty}\sqrt[n]{M_n}.
\end{eqnarray}
Then (\ref{4.14}) also imply that the series and their derivatives
converge in the region $r > \la R_s$.

\section{Interpretation of the multipole moments}
\setcounter{equation}{0}

From (\ref{2.15}) and (\ref{2.20}) we find that, in the canonical
form of the Weyl-Lewis-Papapetrou coordinate system (\ref{1.3}),
the Curzon metric has not `multipole moment' (where it means the
free parameters), but the Schwarzschild metric has infinite ones.
These results are somewhat unnatural and puzzle. Evidently, such
results are caused by the coordinate system, although the
canonical form (\ref{1.3}) is the most convenient one to solve the
metric. So how to extract the understandable information from the
solutions is also an important problem.

Similar to the concepts of point charge and dipole in the
electromagnetism, an ideal explanation for the multipole moments
should be expressed in the forms of some conserved spatial
integrals of the source\cite{intp1}. However, this ideal is
associated with how to define the covariant generalized functions
for nonlinear differential equations, which may have not a general
solution for the higher order moments. A realistic explanation for
these free parameters is to solve them by associating the exterior
solution with the interior one, the results will endow the
parameters with concrete values\cite{appr3}. However, the
influence of the coordinate system still exists. In \cite{intp2},
the authors suggested two ways of carrying out comparison of
approximate and exact solutions: one is calculating the multipole
structure of the Ernst complex potentials for the solutions, and
the second is to generating approximate solutions from exact ones
by expanding the latter in Taylor series with respect to a small
parameter.

To interpret the physical meanings of these free parameters,
introducing a standard ruler may be a convenient choice. As an
approach of first step, we find that the Schwarzschild metric is a
good ruler, because its properties are the simplest and have been
well understood. By comparing the other solutions with this ruler,
we can get some definite and understandable meanings of these free
parameters.

The standard exterior Schwarzschild space-time is described by
\begin{eqnarray}
ds^2= \l(1-\frac {R_s} R\r) dt^2-\l(1-\frac {R_s}
R\r)^{-1}dR^2-R^2d\Th^2-R^2\sin^2\Th d\vf^2. \label{5.1}
\end{eqnarray}
The transformation between (\ref{5.1}) and (\ref{1.3})
reads\cite{exact1}
\begin{eqnarray}
r=\sqrt {R^{2}-2 m R +m^{2} \cos^2\Th}, \qquad \cos \th =\frac { (
R-m ) \cos \Th}{\sqrt {R^{2}-2 m R +m^{2} \cos^2\Th}}, \label{5.2}
\end{eqnarray}
which is valid in the region $R>\frac 1 2 R_s(1+\sin\Th)$. In the
coordinate system $(t,R,\Th,\vf)$, the line element (\ref{1.3})
becomes
\begin{eqnarray}
ds^{2} = U ( dt+W d\vf)^{2} - \wt V [ d R^{2}+ R( R-R_s) d\Th^{2}
] -U^{-1} R( R-R_s)  \sin^2\Th  d\vf^{2}, \label{5.3}
\end{eqnarray}
\begin{eqnarray}
\wt V = \l( 1+ \frac {R_s^{2} \sin^2\Th}{4 R( R- R_s ) }
\r)V.\label{5.4}
\end{eqnarray}
$(t,R,\Th,\vf)$ is also an ACMC-$\infty$ coordinate system.

For the Curzon solution (\ref{2.14}), $(U\to 0,V\to\infty)$
corresponds to the surface
\begin{eqnarray}
R=\frac 1 2 R_s (1+\sin\Th),\qquad \frac 1 2 R_s\le R\le R_s,
\end{eqnarray}
which is an oblate spheroid. This implies the solution is the
analytic extension of the vacuum produced by an ellipsoid.
However, by $\wt V$ in (\ref{5.4}), we find the solutions are only
valid in the region $R>R_s$. Expanding $(U,\wt V)$ in Taylor
series, we get the free parameters for Curzon solution in the form
of Legendre polynomials $P_n(\cos\Th)$,
\begin{eqnarray}
U &=& 1- \frac {R_s}R + \frac 1 {12}{P_2}\l(\frac
{R_s}R\r)^3+\frac 1 {24} P_2 \l(\frac {R_s}R\r)^4+ ({\frac
{1}{56}}P_{{2}}  -{\frac
{3}{ 560}}P_4) \l(\frac {R_s}R\r)^5 + \nn\\
&& ( {\frac {1}{1440}} P_0+{\frac {1}{144}} P_2  -{\frac {1}{160}}
P_4) \l(\frac {R_s}R\r)^6 +O(r^{-7}). \label{5.6}\\
\wt V &=& 1+ \frac {R_s}R+\l(\frac {R_s}R\r)^{2}+ ( P_0-\frac 1
{12} P_{{2}} ) \l(\frac {R_s}R\r)^{3}+ (P_0-{ \frac {29}{168}}
P_{{2}}-\frac 1 {28} P_{{4}} ) \l(\frac {R_s}R\r)^{4}+\nn\\
&& ( P_0-{ \frac {41}{168}} P_{{2}}-{\frac {57}{560}} P_{{4}} )
\l(\frac {R_s}R\r)^{5}+O(r^{-6}).
\end{eqnarray}

For the Kerr solution (\ref{2.16}) and (\ref{2.18}), the multipole
moments in the coordinate system $(t,R,\Th,\vf)$ go as follows
\begin{eqnarray}
U &=& 1- \frac {R_s} R  +\frac 1 4 \al^2 P_{{2}} {\l(\frac {R_s}
R\r)}^{3}+ \al^2( -\frac 1 { 24} P_{{0}}+\frac 1{24} P_{{2}} )
{\l(\frac {R_s} R\r)}^{4}+ \nn \\
&&\al^2 [  ( -\frac 1 {16} \al^{ 2}+{\frac {1}{140}} )
P_{{4}}-\frac 1{40} P_{{0}}+{\frac {1}{56}} P_{ {2}} ] {\l(\frac
{R_s}
R\r)}^{5}+ O(r^{-7})+\nn\\
&& \al^2[ ( {\frac {3}{280}}-{\frac {47}{1120 }} \al^{2} )
P_{{4}}+ (\frac  1{28} \al^{2}+{\frac {1}{168}} ) P_{{2}}+ (
-{\frac {1}{60}}+{\frac {1}{160}} \al^{2} ) P_{{0}} ] {\l(\frac
{R_s} R\r)}^{6}, \label{5.8}
\end{eqnarray}
\begin{eqnarray}
W &=&  \al R_s \sin^2\Th\left\{ \frac 1 2 \frac {R_s} R+\frac 1 2
\l(\frac {R_s} R\r)^{2}+ [ -{\frac {5}{24}} \al^{2}P_{{2}} + (
\frac 1 2-\frac 1
{24} \al^{2} ) P_{{0}} ] \l(\frac {R_s} R\r)^{3}+\r. \nn\\
&& [ -{ \frac {7}{16}} \al^{2}P_{{2}}+ ( \frac 1 2-\frac 1 {16}
\al^{2}
) P_{{0 }} ] \l(\frac {R_s} R\r)^{4}+ O(r^{-7})+ \nn\\
&& [ ( {\frac {9}{160}} \al^{4}-{\frac {3 }{280}} \al^{2} )
P_{{4}}+ ( \frac 1 {32} \al^{4}-{\frac {113}{ 168}} \al^{2} )
P_{{2}}+  \frac 1 2+{\frac {1}{160}} \al^{4}-\frac 1{15} \al^{2}
  ] \l(\frac
{R_s} R\r)^{5}+ \nn\\
&&\l. [  ( {\frac { 79}{448}} \al^{4}-{\frac {19}{560}} \al^{2} )
P_{{4}}+ ( {\frac {113}{1344}} \al^{4}-{\frac {305}{336}} \al^{2}
) P_{{2 }}+  \frac 1 2+\frac 1{48} \al^{4}-{\frac {7}{120}}
\al^{2}  ] \l(\frac {R_s} R\r)^{6} \right\}.~~
\end{eqnarray}
Comparing the dimensionless coefficients of the term
$P_2(\cos\Th)\l(\frac {R_s} R\r)^{3}$ in (\ref{5.6}) and
(\ref{5.8}), namely, $\frac 1 {12}$ and $\frac 1 4 \al^2$,  we get
a definite concept for the relative deformations of each
space-time. The bigger the absolute value of the free
coefficients, the larger the deformation and convection of the
gravitating source. In the viewpoint of series, the Kerr solution
has not any speciality. However, besides $(R_s=2 m, w_1=\frac 1 2
\al=\frac J {m^2})$, whether the other coefficients have some
relations with the general physical concepts is unclear.

The solutions in the metric (\ref{5.3}) can be directly solved
from the following equations
\begin{eqnarray}
R(R-R_s) \pa_R^2u+(2R-R_s) \pa_R u +  {\cot\Th}  \pa_\Th u +
\pa_\Th^2
u +\qquad \qquad\qquad && \nn\\
\frac {2e^u}{\sin^2\Th} [(\pa_R W)^2
+\frac{(\pa_\Th W)^2}{R(R-R_s)}] &=& 0,\label{5.14}\\
 R(R-R_s) \pa_R^2 W - {\cot\Th} \pa_\Th W + \pa_\Th^2 W +
[R(R-R_s) \pa_R u \pa_R W+\pa_\Th u \pa_\Th W] &=&
0.~~\label{5.15}
\end{eqnarray}
The solution is equivalent to (\ref{2.10*})-(\ref{2.11}).

\section{discussion and conclusion}
\setcounter{equation}{0}

The above procedure provides a simple but effective method to
solve the series solution of stationary and asymptotically flat
metric with axisymmetry to any wanted precision. The solution is
identical with two sequences of free dimensionless parameters,
which correspond to the usual multipole moments. The free
parameters are determined by the energy-momentum distribution of
gravitating source. For a wide class of given parameters with
suitable upper bound, the series solution converges in the region
$r>R_s$. For a normal star, we always have $r\gg R_s$, so the
series provides high precise solutions to the stationary metric
with axisymmetry.

To interpret the meanings of free parameters, using dimensionless
form and setting up a standard ruler are meaningful. Only compared
with a simple ruler, we can distinguish the differences between
the solutions, and get a clear concept of the free parameters. For
a given star, to solve the free parameters by associating the
exterior metric with the interior solution, the results will have
concrete physical meanings. However, the matching conditions on
the surface of the star should be carefully discussed. In
\cite{gu}, we find $U\in C^1$ but $V \in C^0$.

\section*{Acknowledgments}
The author is grateful to his supervisor Prof. Ta-Tsien Li and
Prof. Han-Ji Shang for their encouragement.

\newpage
\section*{Appendix}
\setcounter{equation}{0} Substituting (\ref{2.3}) and (\ref{2.4})
into the original equation (\ref{2.1}) and (\ref{2.2}), and
expanding them in series, we get the relations among the
parameters $(a_{kn},b_{kn})$ as follows.
\begin{eqnarray}
0 &=& 22 a_{20} \l(\frac {R_s}r\r)^2 + (-2 a_{3,1} + 6 a_{30})
\l(\frac {R_s}r\r)^3 +\nn \\
&&[ (6 a_{41}+ 12 b_{11}^{2} ) \cos2\th + (-2 a_{41}+12
a_{40} ) +20 b_{11}^{2}  ] \l(\frac {R_s}r\r)^4 +\nn \\
&& [ ( ( -8 a_{{5,2}}+14 a_{{5,1}} ) +4 b_{{1,1}} ( 3 a
_{{1,0}}b_{{1,1}}+4 b_{{2,1}} ) ) \cos 2\th  +\nn \\
&&( -4 a_{{5,2}}-2 a_{{5,1}}+20 a_{{5,0}} ) +4
 b_{{1,1}} ( 5 a_{{1,0}}b_{{1,1}}+12 b_{{2,1}} )
 ] \l(\frac {R_s}r\r)^5 +O(r^{-6}),\label{6.1}\\
0 &=& \l\{ (8 b_{21} + 2 a_{10}b_{11}) \l(\frac {R_s}r\r)^2 + [ (
16 b_{32}+20 b_{31} ) +4 a_{20}b_{11}+4 a_{10}b_{21} ] \l(\frac
{R_s}r\r)^3 +\r.\nn \\
&& [  (32 b_{{4,2}} -2 a_{{3,1}}b_{{1,1}}+12 a_{{1,0}}b_{{3,2}} )
\cos  2 \theta
 +( 48 b_{{4,2}}+36 b_{{4,1}}) + \nn \\
 && 6 b_{{1,1}}a_
{{3,0}}-8 a_{{3,1}}b_{{1,1}}+12 a_{{1,0}}b_{{3,2}}+8
a_{{2,0}}b_{{2 ,1}}+6 a_{{1,0}}b_{{3,1}} ] \l(\frac
{R_s}r\r)^4+ \nn \\
&&[  \left(  ( 48 b_{{5,3}}+72 b_{{5,2}} ) +16 a_{{
1,0}}b_{{4,2}}+4 a_{{3,1}}b_{{2,1}}+24 a_{{2,0}}b_{{3,2}} \right)
\cos 2 \theta +\nn \\
&& ( 72 b_{{5,3}}+88 b_{{5,2}}+56
 b_{{5,1}} )
+8 b_{{1,1}}a_{{4,0}}+16 a_{{1,0}}b_{{4,2}}-8
a_{{3,1}}b_{{2,1}}+24 a_{{2,0}}b_{{3,2}}+\nn \\
&&\l. 12 a_{{2,0}}b_{{3,1}}+12 b _{{2,1}}a_{{3,0}}-8
a_{{4,1}}b_{{1,1}}+8 a_{{1,0}}b_{{4,1}} ] \l(\frac {R_s}r\r)^5
+O(r^{-6}) \r\}R_s\sin^2\th. \label{6.2}
\end{eqnarray}
The the solutions of the parameters are given by the following
recursive relations
\begin{eqnarray}
&& a_{10} = -2,~~~~ a_{20} = 0,~~~  a_{31} = m_1,~~~ a_{30} =
\frac 1 3 m_1,~~~ a_{41} = -\frac 1 2 w_1^2,~~~ a_{40} =
-\frac 1 2 w_1^2, \nn\\
&&a_{52} = m_2, ~~~ a_{51} = \frac 2 7 w_1^2+\frac 4 7 m_2, ~~~
a_{50} =
\frac 8 {35}  w_1^2+\frac 9 {35} m_2, \nn\\
&& a_{62}=-\frac 1 2 w_{1}w_{2},~~~a_{61}=-\frac 1 {10}
{w_1}^{2}-\frac 4 5 w_{1}w_{2},~~~
a_{60}=-\frac 1 {10} {w_1}^{2}-\frac 3 {10} w_1 w_2, ~~\cdots \nn\\
&&b_{11} = -\frac 1 2 w_1, ~~~ b_{21} = -\frac 1 4 w_1, ~~~ b_{32}
= -\frac 1 4 w_2, ~~~ b_{31}
= \frac 1 5 w_2-\frac 1 {10} w_1, \nn \\
&&b_{42} = -\frac 1 {32} m_1 w_1-\frac 3 {16} w_2,  ~~~b_{41} =
-\frac 1 {30} w_1+\frac 3 {20} w_2-\frac 1 {24} m_1 w_1,\label{6.3} \\
&&b_{53} = -\frac 1 6 w_3,~~~ b_{52} = -\frac 1 {12} w_2+\frac 1 9
w_3, ~~~b_{51} = -\frac 1 {21} m_1 w_1-\frac 1 {105} w_1+\frac
1{15} w_2+\frac 5 {126} w_3\nn\\
&& b_{63}=-{\frac {5}{36}} w_3-\frac 1{16} w_{1}m_{2}+{\frac
{1}{96 }} m_{1}w_{2} ,\nn\\
&& b_{62}=-{\frac {1}{168}} {w_1}^{3}-{\frac {3}{56}} w_1m_{
{2}}+{\frac {1}{240}} m_1w_2+{\frac {1}{120}} m_1w_1
-\frac 1 {36} w_2+{\frac {5}{54}} w_3\nn\\
&& b_{61}={\frac {11}{480}} m_1w_2-{\frac {1}{420}} {w_1
}^{3}+{\frac {51}{560}} w_1m_2-\frac 1 {35} m_1w_1+\frac 1 {45}
w_{ {2}}+{\frac {25}{756}} w_3-{\frac {1}{420}} w_1, ~~\cdots \nn
\end{eqnarray}

The solutions to (\ref{5.14}) and (\ref{5.15}) in the metric
(\ref{5.3}) are given by
\begin{eqnarray} U&=&1-\frac
{R_s}R+ ( \frac 1 2 M_1 \cos 2\Th +\frac 1 6 M_1 )
\l(\frac {R_s}R\r)^3+\nn\\
&& [( \frac 1 4 M_1 -\frac 1 4 w_1^{2} ) \cos 2 \Th +\frac 1 {12}
M_1-\frac 1 4 w_1^2 ] \l(\frac {R_s}R\r)^4+O(r^{-5}), \label{5.10}\\
\wt V &=& 1+\frac {R_s}R + \l(\frac {R_s}R\r)^{2}+ ( 1-\frac 1 2
M_1
\cos 2 \Th - \frac 1 6  M_1  ) \l(\frac {R_s}R\r)^{3}+ O(r^{-5})+ \l(1 -{\frac {37}{96}} M_1 +\r.\nn\\
&& \l.  {\frac {11}{64} } {w_1 }^{2}+ ( -{\frac {5}{32}} M_1 +{
\frac {9}{64}} {w_1 }^{2} ) \cos 4 \Th + ( -{\frac {9}{8}} M_1
+\frac 3{16} {w_1 }^{2} ) \cos
 2 \Th \r)  \l(\frac {R_s}R\r)^{4},\\
{W} &=& { R_s}\sin^2\Th \left\{ w_1 \frac {R_s}R +w_1 \l(\frac
{R_s}R\r)^{2}+O(r^{-5}) + \r.\nn\\
&&[  ( w_2 -\frac 3{16} w_1 ) \cos  2 \Th +\frac 3 5 w_2 +{\frac
{71}{80}}
 w_1 ] \l(\frac {R_s}R\r)^{3}+  \nn\\
 &&\l. [  ( -{\frac
{27}{64}} w_1 +\frac 1 8 M_1 w_1 +\frac 9 4 w_2 ) \cos 2 \Th
+{\frac {239}{320}} w_1 +{\frac {5}{24}} M_1 w_1 +{\frac {
27}{20}} w_2 ] \l(\frac {R_s}R\r)^{4} \right\},~~ \label{5.12}
\end{eqnarray}
where
\begin{eqnarray}
M_1 =m_1 + \frac 1 8 ,\quad  M_2 = m_2+{\frac { 7}{2^{9}}},\quad
M_3 =m_3 +{\frac { 3\cdot 11 }{ 2^{14}}},~~\cdots
\end{eqnarray}
are somewhat `pure mass' multipole moments deducted the influence
of the coordinates (see (\ref{2.20})). In the case of
Schwarzschild metric, we have $(M_k=w_k = 0,\forall k)$.

\end{document}